\documentclass[aps,pra,preprint,showpacs]{revtex4}
\usepackage{graphicx}
\usepackage{dcolumn}                 
\usepackage{times}
\usepackage{amsmath}

\newcommand*{\cent}[1]{\multicolumn{1}{c}{$#1$}}
\newcolumntype{w}[1]{D{.}{.}{#1}}
\newcolumntype{.}{D{x}{}{-1}}

\begin{document}
\preprint{Version 1.0}

\title{Accurate Born-Oppenheimer potential for HeH$^+$}

\author{Krzysztof Pachucki}
\affiliation{Faculty of Physics, University of Warsaw,
             Ho\.{z}a 69, 00-681 Warsaw, Poland}

\begin{abstract}
We demonstrate high accuracy calculations for the HeH$^+$ molecule
using newly developed analytic formulas for two-center two-electron 
integrals with exponential functions. The Born-Oppenheimer potential 
for the ground electronic $^1\Sigma^+$ state is obtained
in the range of 0.1 -- 60 au with precision of about $10^{-12}$ au.  
As an example at the equilibrium distance $r=1.463\,283$ au 
the Born-Oppenheimer potential amounts to $-2.978\,708\,310\,771(1)$.  
Obtained results lay the ground for theoretical predictions in HeH$^+$
with spectroscopic precision.
\end{abstract}

\maketitle

\section{Introduction}
HeH$^+$ molecule is a simple two electron molecule consisting of 
the $\alpha$ particle and of the proton as nuclei. In the ground electronic $^1\Sigma^+$ state,
both electrons are mostly centered around $\alpha$ nucleus with the proton
distance from $\alpha$ being about $r=1.46$ au. 
The first accurate variational calculations of the Born-Oppenheimer (BO) potential of HeH$^+$
reported by Wolniewicz \cite{wolniewicz_65}, Ko\l os and Peek \cite{kolos_peek},
Ko\l os \cite{kolos_76}, were shortly afterwards refined by Bishop and Cheung \cite{bishop_79}.
In these calculations authors represented the electronic wave function in terms of
exponential functions times various polynomials in
interparticle distances with general nonlinear parameters 
(Ko\l os-Wolniewicz functions) or with some parameters fixed (James-Coolidge functions).
Accuracy achieved in calculations by Bishop and Cheung was about $4\cdot10^{-6}$ au.
More accurate result $\sim 10^{-8}$ au, at the equilibrium distance $r=1.46$ au,  
have been obtained by Cencek {\em et al}. \cite{cencek_95}
using explicitly correlated Gaussian functions.
Even if the whole BO potential is known with their accuracy ($\sim$ 0.02 cm$^{-1}$),
it will be insufficient in comparison with the most accurate 
($\sim 10^{-5}$ cm$^{-1}$)
measurements of rovibrational transition frequencies \cite{hehp_exp1}. 
More recently Adamowicz {\it et al}. \cite{adam1, adam2,adam3} performed direct 
nonadiabatic calculations for HeH$^+$ including relativistic corrections, 
however, only for purely vibrational states and without estimation of accuracy.   

In this work we use asymptotically correct generalized Heitler-London functions, which are 
the product of an exponential atomic He function times an arbitrary polynomial in interparticle distances.
Thanks to analytic formulas for the resulting two-center two-electron integrals \cite{rec_h2},
we perform calculations with a large number $\sim 20\,000$ of basis functions. The resulting accuracy
of about $10^{-12}$ au, or better (for large distances) is 4 orders of magnitude
improvement with respect to previous values. The accuracy of BO potential is not directly transformed
into accuracy of rovibrational energy levels. It is because Born-Oppenheimer energies 
should be supplemented by adiabatic, nonadiabatic, relativistic and QED corrections.
As it has been recently demonstrated for H$_2$ and isotopomers \cite{h2},
the perturbative treatment in the electron-nuclei mass ratio up to $O(m/M)^2$
and the fine structure constant $\alpha$ up to $O(\alpha)^3$
can provide rovibrational transitions with accuracy reaching $0.000\,1$ cm $^{-1}$. 
Analogous calculations can be  performed for HeH$^+$, and results obtained here are the first
step  for spectroscopically accurate theoretical predictions. 

\section{ECA basis set}
We represent the ground state nonrelativistic electronic wave function 
in terms of generalized Heitler-London functions, which are 
the product of the atomic He exponential function with
an arbitrary polynomial in all interparticle distances, 
\begin{eqnarray}
\phi &=& \sum_{\{n\}}\,c_{\{n\}}(1+P_{12})\,e^{-\alpha\,(r_{1A}+r_{2A})}
\,r_{12}^{n_1}\,r_{1A}^{n_2}\,r_{1B}^{n_3}\,r_{2A}^{n_4}\,r_{2B}^{n_5}
\end{eqnarray}
such that
\begin{equation}
\sum_{i=1}^5 n_i \leq \Omega,
\end{equation}
for some integer $\Omega$,
and call them the explicitly correlated asymptotic (ECA) basis.
In the above $c_{\{n\}}$ are linear coefficients, $\{n\}$ represents
a set of 5 numbers $\{n_1, n_2, n_2, n_4, n_5\}$, $P_{12}$
is the operator which replaces $\vec r_1$ with $\vec r_2$,
indices $(A, B)$ denote nuclei, and $(1, 2)$ denote electrons.
Matrix elements of the nonrelativistic Hamiltonian
can be expressed in terms of integrals of the form 
\begin{equation}
f(n_1,n_2,n_3,n_4,n_5;r,\beta) = \int \frac{d^3 r_1}{4\,\pi}\,\int \frac{d^3 r_2}{4\,\pi}\,
\frac{r}{r_{12}^{1-n_1}}\,
\frac{e^{-\beta\,r_{1A}}}{r_{1A}^{1-n_2}}\,
\frac{1}{r_{1B}^{1-n_3}}\,
\frac{e^{-\beta\,r_{2A}}}{r_{2A}^{1-n_4}}\,
\frac{1}{r_{2B}^{1-n_5}}.
\end{equation}
with nonnegative integers $n_i$.
These integrals are calculated as follows.
When all $n_i=0$ the so called master integral, which can be derived from 
a general expression in Ref. \cite{rec_h2}, is
\begin{eqnarray}
f(r,\beta) &=& \int \frac{d^3 r_1}{4\,\pi}\,\int \frac{d^3 r_2}{4\,\pi}\,
\frac{e^{-\beta\,r_{1A}}}{r_{1A}}\,\frac{1}{r_{1B}}\,\frac{e^{-\beta\,r_{2A}}}{r_{2A}}\,
\frac{1}{r_{2B}}\,\frac{r}{r_{12}} \nonumber \\ &=&
 \frac{1}{\beta^2}\,\int_0^{\beta\,r} dx\,F(x) = 
-\frac{1}{\beta^2}\,\int^\infty_{\beta\,r} dx\,F(x) \label{04}
\end{eqnarray}
with
\begin{eqnarray}
F(x) &=& \frac{1}{x}\,\biggl[\frac{e^{2\,x}}{2}\,{\rm Ei}(-2\,x) -\frac{e^{-2\,x}}{2}\,{\rm Ei}(2\,x)
          - {\rm Ei}(-x) + e^{-2\,x}\,{\rm Ei}(x)\biggr], \label{05}
\end{eqnarray}
where Ei is the exponential integral function.
The one-dimensional integral in Eq. (\ref{04})
is calculated numerically using the adapted Gaussian integration.
With 120 integration points one achieves 64 significant digits for all the distances.
All $f$-integrals with  higher powers of electron distances can be obtained from recursion
relations, which were derived in Ref. \cite{rec_h2}. Since  
\begin{equation}
f(n_1,n_2,n_3,n_4,n_5;r,\beta) = 
f(n_1,n_2,n_3,n_4,n_5,1,\beta\,r)\,r^{2+n_1+n_2+n_3+n_4+n_5}
\end{equation}
we present, as an example, formulas at $r=1$ for all integrals with $\sum_i n_i\leq 2$
\begin{eqnarray}
 f(0, 0, 0, 0, 0) &=& f \\
 f(0, 0, 0, 0, 1) &=& \frac{1}{2\,\beta^3} + \frac{e^{-2\,\beta}}{2\,\beta^3} 
                     - \frac{e^{-\beta}}{\beta^3} + 
     \frac{[{\rm Ei}(-2\,\beta) - {\rm Ei}(-\beta)]}{\beta^2} \nonumber \\
 f(0, 0, 0, 1, 0) &=& - \frac{F}{2\,\beta^3} + \frac{f}{\beta}\nonumber \\
 f(1, 0, 0, 0, 0) &=& \frac{1}{\beta^4} + \frac{e^{-2\,\beta}}{\beta^4} 
                        - \frac{2\,e^{-\beta}}{\beta^4}\nonumber \\
 f(0, 0, 0, 0, 2) &=& \frac{1}{2\,\beta^3} - \frac{e^{-\beta}}{\beta^3} 
                      - \frac{{\rm Ei}(-\beta)}{\beta^2} + \frac{f}{\beta^2}\nonumber \\
 f(0, 0, 0, 1, 1) &=& \frac{3}{4\,\beta^4} + \frac{5\,e^{-2\,\beta}}{4\,\beta^4} 
                      - \frac{2\,e^{-\beta}}{\beta^4} 
                      + \frac{2\,[{\rm Ei}(-2\,\beta) - {\rm Ei}(-\beta)]}{\beta^3}\nonumber \\
 f(0, 0, 0, 2, 0) &=& -\frac{e^{-\beta}}{2\,\beta^3} + \frac{e^{2\,\beta}\,{\rm Ei}(-2\,\beta)}{2\,\beta^3}
                      - \frac{(1+\beta)\,{\rm Ei}(-\beta)}{2\,\beta^3}
                      -\frac{(5 + 2\,\beta)\,F}{4\,\beta^4} 
                      + \frac{2\,f}{\beta^2}\nonumber \\
 f(0, 0, 1, 0, 1) &=& \frac{1}{2\,\beta^3}\nonumber \\
 f(0, 0, 1, 1, 0) &=& \frac{3}{4\,\beta^4} + \frac{e^{-2\,\beta}}{4\,\beta^4} - \frac{e^{-\beta}}{\beta^4}\nonumber \\
 f(0, 1, 0, 1, 0) &=& \frac{e^{-\beta}}{2\,\beta^3} + \frac{e^{2\,\beta}\,{\rm Ei}(-2\,\beta)}{2\,\beta^3}
             + \frac{(1-\beta)\,{\rm Ei}(-\beta)}{2\,\beta^3}
             - \frac{(5 + 2\,\beta)\,F}{4\,\beta^4}
             + \frac{f}{\beta^2}\nonumber \\
 f(1, 0, 0, 0, 1) &=& \frac{1}{\beta^4} - \frac{e^{-\beta}}{\beta^4}\nonumber \\
 f(1, 0, 0, 1, 0) &=& \frac{2}{\beta^5} + \frac{(2 + \beta)\,e^{-2\,\beta}}{\beta^5} 
                      - \frac{(4 + \beta)\,e^{-\beta}}{\beta^5}\nonumber \\
 f(2, 0, 0, 0, 0) &=& \frac{2}{\beta^5} - \frac{2\,(1 + \beta)\,e^{-\beta}}{\beta^5} 
                      - \frac{2\,e^{2\,\beta}\,{\rm Ei}(-2\,\beta)}{\beta^5} + 
     \frac{2\,(1 + \beta)\,{\rm Ei}(-\beta)}{\beta^5} \nonumber \\ && 
    +\frac{(1 + 2\,\beta)\,F}{\beta^6} + \frac{2\,f}{\beta^2}
\end{eqnarray}
where $f = f(1,\beta)$ and $F=F(\beta)$.
Other $f$-integrals from the same shell can be obtained by using the symmetry 
$f(n_1,n_2,n_3,n_4,n_5) = f(n_1,n_4,n_5,n_2,n_3)$.
Integrals with higher powers $n_i$ are of analogous form,
they are  all  linear combinations of $f$, $F$, Ei, Exp, and identity functions
with coefficients being polynomials in $1/\beta$.
Using a computer symbolic program, 
we have generated a table of integrals with $\sum_i n_i \leq 37$,
which corresponds to a maximum value of $\Omega=16$. 

\section{Numerical results}
The matrix elements of the nonrelativistic Hamiltonian  between ECA
functions are obtained as described by Ko\l os and Roothan in \cite{kolroth}. 
The resulting expression is a linear combination of various ECA integrals 
which are calculated by using analytic formulas as presented in the previous
section. Their evaluation is fast and accurate, thus suitable for
calculations involving large number of basis functions.

Eigenvalues of the Hamiltonian matrix are obtained 
by inverse iteration method for various length of basis sets ($\Omega=7-16$).
These eigenvalues, obtained at 132 internuclear distances 
ranging from 0.1 au to 60 au, form a BO potential.
ECA functions are especially suitable for large distances,
as they include functions for the helium atom alone.
As a result, the atomic He energy is obtained with an accuracy of about $10^{-15}$
for all the distances, in fact much better than that for HeH$^+$. 
In order to improve numerical accuracy for HeH$^+$,
we used a triple basis set $(\Omega, \Omega-2,\Omega-4)$
each one with its own optimized nonlinear parameter $\alpha$. 
Numerical calculations are performed for $r\leq 12$ au
using the quadruple precision, and for $r>12$ au
using the octuple precision arithmetics.
In order to check numerics, we repeated calculations
around the equilibrium distance $r=1.463\,283$ au  by using 
James-Coolidge basis set, for which analytic formulas have been 
developed in Ref. \cite{bo_h2}.
This basis has a slower numerical convergence but is numerically more stable,
so we could use even larger number $N=27\,334$ of basis functions.
Obtained results have similar accuracy and are in perfect agreement with 
that obtained with ECA functions.

The most accurate variational result reported at the distance $r=1.46$ au 
is compared in Table \ref{table2} to all the previous results obtained so far in the literature. 
\begin{table}[htb]\renewcommand{\arraystretch}{0.85}
\caption{\label{table2} Variational Born-Oppenheimer potential for the HeH+ molecule at $r=1.46$ au}
\begin{ruledtabular}
  \begin{tabular}{l.}
 Authors                                                         & {\rm energy[au]}            \\ \hline\\[-5pt]
 1965\; L. Wolniewicz \cite{wolniewicz_65}                       & -2.978\,x666\,7             \\
 1994\; W. Ko\l os and L. Peek \cite{kolos_peek}                 & -2.978\,x689\,06            \\
 1979\; D.M. Bishop and L.M. Cheung \cite{bishop_79}             & -2.978\,x702\,62            \\
 2006\; W. Cencek, J. Komasa and J. Rychlewski \cite{cencek_95}  & -2.978\,x706\,591           \\
 2012\; this work                                                & -2.978\,x706\,600\,341  
  \end{tabular}
\end{ruledtabular}
\end{table}

In performing extrapolation to a complete basis set $(\Omega\rightarrow\infty)$,  
similarly to the previous H$_2$ case \cite{bo_h2},
we observe the exponential $e^{-\beta\,\Omega}$
convergence, in other words the $\log$ of differences
in energies for subsequent values of $\Omega$ fits well to the linear
function. This makes extrapolation to infinity quite simple.
Extrapolated results for the whole BO potential
curve in the range $0.1 - 60$ au are presented in Table \ref{table3}.
{\squeezetable
\begin{table}[htb]\renewcommand{\arraystretch}{0.85}
\caption{\label{table3} Numerical values for BO potential at different internuclear distance $r$,
                        shifted by the He ground state energy $E=-2.903\,724\,377\,034\,120$ au. 
                        Results are obtained by extrapolation to the complete set of basis functions,
                        $r_{\rm min} = 1.463\,283$ au.}
\begin{ruledtabular}
\begin{tabular*}
{1.00\textwidth}{w{0.1}w{1.14}@{\hfill}w{1.0}w{1.15}@{\hfill}w{1.0}w{1.15} @{\hfill}w{2.0}w{1.15}}
\cent{r/\mathrm{au}} & \cent{E} &\cent{r/\mathrm{au}} & \cent{E} &\cent{r/\mathrm{au}} & \cent{E} &\cent{r/\mathrm{au}} & \cent{E} \\
\hline
    0.10& 15.765\,275\,621\,061(3)    &1.60 & -0.072\,566\,613\,147(4)    &3.25 &- 0.010\,505\,993\,345\,15(2) & 5.80 & -0.000\,684\,274\,833\,49(5)\\
    0.20&  6.050\,585\,777\,587\,9(4) &1.65 & -0.070\,782\,049\,442(4)    &3.30 & -0.009\,805\,291\,916\,24(2) & 5.90 & -0.000\,634\,959\,821\,45(7)\\
    0.30&  3.035\,104\,004\,312\,9(1) &1.70 & -0.068\,681\,986\,181(4)    &3.35 & -0.009\,153\,886\,966\,24(2) & 6.00 & -0.000\,590\,184\,825\,0(1) \\
    0.40&  1.676\,798\,340\,477\,6(1) &1.75 & -0.066\,346\,472\,331(4)    &3.40 & -0.008\,548\,577\,089\,50(2) & 6.20 & -0.000\,512\,278\,183\,5(1) \\
    0.50&  0.961\,224\,244\,974\,8(1) &1.80 & -0.063\,841\,750\,276(3)    &3.45 & -0.007\,986\,304\,940\,94(2) & 6.40 & -0.000\,447\,216\,535\,6(6) \\
    0.60&  0.551\,425\,923\,210\,9(2) &1.85 & -0.061\,222\,573\,648(3)    &3.50 & -0.007\,464\,161\,626\,98(2) & 6.60 & -0.000\,392\,456\,561\,3(6) \\
    0.70&  0.305\,137\,807\,453\,6(3) &1.90 & -0.058\,534\,109\,374(3)    &3.55 & -0.006\,979\,388\,833\,37(2) & 6.80 & -0.000\,346\,039\,248\,5(6) \\
    0.80&  0.153\,061\,340\,049\,4(5) &1.95 & -0.055\,813\,503\,899(3)    &3.60 & -0.006\,529\,379\,042\,23(2) & 7.00 & -0.000\,306\,439\,293\,9(6) \\
    0.90&  0.058\,077\,362\,203\,3(8) &2.00 & -0.053\,091\,177\,182(2)    &3.65 & -0.006\,111\,674\,157\,71(2) & 7.20 & -0.000\,272\,456\,681\,9(6) \\
    0.95&  0.024\,964\,448\,474\,0(9) &2.05 & -0.050\,391\,895\,330(2)    &3.70 & -0.005\,723\,962\,825\,47(2) & 7.40 & -0.000\,243\,137\,960\,03(5) \\
    1.00& -0.001\,080\,054\,475(1)    &2.10 & -0.047\,735\,662\,718(2)    &3.75 & -0.005\,364\,076\,701\,23(2) & 7.60 & -0.000\,217\,718\,542\,45(4) \\
    1.05& -0.021\,440\,483\,674(2)    &2.15 & -0.045\,138\,466\,546(2)    &3.80 & -0.005\,029\,985\,892\,14(2) & 7.80 & -0.000\,195\,580\,003\,55(3) \\
    1.10& -0.037\,213\,297\,730(2)    &2.20 & -0.042\,612\,900\,482(1)    &3.85 & -0.004\,719\,793\,768\,31(2) & 8.00 & -0.000\,176\,218\,146\,92(3) \\
    1.15& -0.049\,273\,330\,916(2)    &2.25 & -0.040\,168\,689\,007(1)    &3.90 & -0.004\,431\,731\,314\,03(2) & 8.50 & -0.000\,137\,416\,959\,04(5) \\
    1.20& -0.058\,323\,089\,351(2)    &2.30 & -0.037\,813\,130\,017\,8(9) &3.95 & -0.004\,164\,151\,165\,85(2) & 9.00 & -0.000\,108\,799\,429\,7(1) \\
    1.25& -0.064\,929\,882\,196(3)    &2.35 & -0.035\,551\,469\,970\,6(7) &4.00 & -0.003\,915\,521\,460\,65(2) & 9.50 & -0.000\,087\,296\,341\,4(7) \\
    1.30& -0.069\,554\,101\,696(3)    &2.40 & -0.033\,387\,223\,169\,6(6) &4.10 & -0.003\,469\,526\,010\,47(2) & 10.0 & -0.000\,070\,874\,764(5)    \\
    1.35& -0.072\,570\,979\,391(3)    &2.45 & -0.031\,322\,444\,662\,9(5) &4.20 & -0.003\,083\,563\,658\,79(2) & 10.5 & -0.000\,058\,152\,803(6)    \\
    1.38& -0.073\,739\,700\,599(4)    &2.50 & -0.029\,357\,964\,421\,4(4) &4.30 & -0.002\,748\,907\,761\,76(2) & 11.0 & -0.000\,048\,170\,045(4)    \\
    1.40& -0.074\,287\,476\,582(4)    &2.55 & -0.027\,493\,589\,050\,7(3) &4.40 & -0.002\,458\,102\,556\,14(2) & 11.5 & -0.000\,040\,245\,989\,5(9) \\
    1.41& -0.074\,497\,742\,079(2)    &2.60 & -0.025\,728\,276\,107\,6(2) &4.50 & -0.002\,204\,793\,867\,58(2) & 12.0 & -0.000\,033\,890\,131\,5(2) \\
    1.42& -0.074\,667\,994\,434(2)    &2.65 & -0.024\,060\,285\,148\,3(2) &4.60 & -0.001\,983\,576\,228\,38(2) & 13.0 & -0.000\,024\,539\,525\,11(1)\\
    1.43& -0.074\,799\,949\,340(2)    &2.70 & -0.022\,487\,308\,864\,0(1) &4.70 & -0.001\,789\,856\,934\,04(3) & 14.0 & -0.000\,018\,206\,960\,42(1)\\
    1.44& -0.074\,895\,257\,253(1)    &2.75 & -0.021\,006\,587\,043\,5(1) &4.80 & -0.001\,619\,736\,634\,04(3) & 15.0 & -0.000\,013\,793\,858\,65(1)\\
    1.45& -0.074\,955\,506\,003(1)    &2.80 & -0.019\,615\,005\,609\,25(9)&4.90 & -0.001\,469\,905\,474\,75(3) & 16.0 & -0.000\,010\,641\,724\,04(1)\\
    1.46& -0.074\,982\,223\,308(1)    &2.85 & -0.018\,309\,182\,582\,72(7)&5.00 & -0.001\,337\,553\,485\,43(4) & 17.0 & -0.000\,008\,341\,461\,47(1)\\
    r.\!_{\rm \!min}
        & -0.074\,983\,933\,737(1)    &2.90 & -0.017\,085\,542\,525\,64(6)&5.10 & -0.001\,220\,293\,751\,56(4) & 18.0 & -0.000\,006\,630\,920\,31(1)\\
    1.47& -0.074\,976\,879\,155(1)    &2.95 & -0.015\,940\,380\,764\,26(5)&5.20 & -0.001\,116\,096\,895\,45(4) & 19.0 & -0.000\,005\,337\,497\,57(1)\\
    1.48& -0.074\,940\,888\,099(1)    &3.00 & -0.014\,869\,918\,514\,72(4)&5.30 & -0.001\,023\,235\,434\,23(4) & 20.0 & -0.000\,004\,344\,797\,44(1)\\
    1.49& -0.074\,875\,611\,446(1)    &3.05 & -0.013\,870\,349\,882\,66(3)&5.40 & -0.000\,940\,236\,685\,33(4) & 30.0 & -0.000\,000\,855\,679\,62(1)\\
    1.50& -0.074\,782\,359\,346(1)    &3.10 & -0.012\,937\,881\,595\,48(3)&5.50 & -0.000\,865\,843\,012\,73(5) & 40.0 & -0.000\,000\,270\,476\,79(1)\\
    1.52& -0.074\,516\,925\,580(3)    &3.15 & -0.012\,068\,766\,236\,41(3)&5.60 & -0.000\,798\,978\,332\,42(5) & 50.0 & -0.000\,000\,110\,738\,51(1)\\
    1.55& -0.073\,938\,987\,047(4)    &3.20 & -0.011\,259\,329\,676\,92(2)&5.70 & -0.000\,738\,719\,940\,4(4)  & 60.0 & -0.000\,000\,053\,391\,46(1)
\end{tabular*}
\end{ruledtabular}
\end{table}}

\section{Summary}
We have demonstrated applications of analytic formulas for
two-center two-electron exponential integrals in high precision calculations
of Born-Oppenheimer potential for the ground electronic state
of HeH$^+$, similarly to previous calculations for H$_2$ \cite{bo_h2}. 
The use of ECA basis 
with as much as 20~000 functions provided energies with precision 
of about $10^{-12} - 10^{-14}$ au for internuclear distances up to 60 au.

The extension of this approach to calculations of excited states of diatomic molecules
such as H$_2$ and HeH$^+$ is not difficult. As long as, the trial function includes
at most two different nonlinear parameters (in the exponent), the analytic formulas
for corresponding integrals are not exceedingly large. 
Such two parameters exponential functions can represent well an arbitrary electronic state
of a diatomic molecule. In fact, we aim to develop  
a general code for calculation of BO potential 
of the diatomic molecule, using quad and the arbitrary precision arithmetics. The extension to
adiabatic, nonadiabatic and relativistic corrections is more problematic.
For their evaluation one needs integrals with inverse powers of
interparticle distances, for which formulas can be quite complicated.
We think that their derivation for James-Coolidge basis is within the reach,
therefore for small internuclear distances, where this basis works quite well,
all relevant corrections, including yet unknown $\alpha^4$ QED correction
can be calculated to high precision, thus improving theoretical predictions
for H$_2$ and other two-electron diatomic molecules to about $10^{-6}$ cm$^{-1}$.

\end{document}